\documentclass[prb,twocolumn,showpacs,showkeys,preprintnumbers,amsmath,amssymb]{revtex4}
\usepackage{graphicx}
\usepackage[T1]{fontenc}
\usepackage{amssymb,amsmath}
\usepackage{dcolumn}% Align table columns on decimal point
\usepackage{bm}% bold math 

\begin{document}

\title{Exact results in a slave boson saddle point approach \\
for a strongly correlated electron model}
\author{Raymond Fr\'esard$^1$ and Thilo Kopp$^2$ }
\affiliation{$^1$Laboratoire CRISMAT UMR CNRS--ENSICAEN 6508, and IRMA,
  FR3095, Caen, France\\ 
  $^2$Center for Electronic Correlations and Magnetism, Institute of Physics, 
                      Universit\"at Augsburg, D-86135 Augsburg, Germany} 
%\author{Thilo Kopp}
%\affiliation{Center for Electronic Correlations and Magnetism, Institute of Physics, 
%                      Universit\"at Augsburg, D-86135 Augsburg, Germany}
\date{\today}% It is always \today, today,
             %  but any date may be explicitly specified

\begin{abstract}
We revisit the Kotliar-Ruckenstein (KR) slave boson saddle point evaluation
for a two-site correlated electron model. As the model can be solved
analytically, it is possible to compare the KR saddle point results to the
exact many particle levels. The considered two site cluster mimics an
infinite-$U$ single-impurity Anderson model with a nearest neighbor Coulomb
interaction: one site is strongly correlated with an infinite local Coulomb
repulsion which hybridizes with the second site, on which the local Coulomb
repulsion vanishes. Making use of the flexibility of the representation we
introduce appropriate weight factors in the KR saddle point scheme. Ground
state and all excitation levels agree with the exact diagonalization results.  
Thermodynamics and correlation functions may be recovered in a suitably
renormalized  saddle point evaluation.

\end{abstract}
\maketitle

The unusual properties of the high temperature superconductors or the colossal 
magnetoresistance and orbital ordering in transition metal oxides
\cite{reviews} are calling for new techniques and concepts to describe these
phenomena. A formalism, devised to deal specifically with strongly correlated
electron systems, is provided by the slave boson approach pioneered by Barnes
\cite{BAR76} for the single impurity Anderson model (SIAM), and further
extended to the Hubbard model by Kotliar and Ruckenstein. \cite{KOT86} 
The approach implements a local decomposition of electronic excitations into
charge and spin components which is achieved by the introduction of composite
operators for all local (on-site) excitations. The composite operators
separate into canonical bosonic and fermionic operators where, in the Barnes
approach, the charge (spin) degrees of freedom are represented by bosonic
(fermionic) operators. However the latter are enslaved in the sense
that their respective number operators have to fulfill a local constraint. 
The original idea was to decouple spin and charge degrees of freedom. In the
Kotliar-Ruckenstein (KR) scheme a bosonic mode is attributed to each type of
excitation which allows to study the correlated system in a saddle point
approximation for all degrees of freedom.\cite{KOT86, FRE92, HAS97} This
latter approach has been  impressively successful when compared to numerical
simulations: ground state energies~\cite{FRE91} and charge structure factors
show excellent agreement.\cite{ZIM97}

The slave boson approach has several intriguing properties. Firstly, the
approach is exact in the large degeneracy limit.\cite{FRE92,FLO02} Moreover,
the paramagnetic mean-field solution reproduces the Gutzwiller approximation
in the KR representation.\cite{KOT86} It obeys a variational principle in the
limit of large spatial dimensions where the Gutzwiller approximation and the
Gutzwiller wave function are identical.\cite{MET89} These formal properties
signify that the approach captures characteristic features of strongly
correlated electrons as the suppression of the quasiparticle weight and the
Mott-Hubbard/Brinkman-Rice transition \cite{BRI70} to an insulating state at
half filling with increasing on-site Coulomb interaction. 

Secondly, as the fields are canonical within this approach, long range
correlations are more amenable to analyses. Moreover, one can easily introduce
a long range Coulomb interaction. In a radial gauge representation of the
slave bosons, long range Coulomb interactions can even be  cast into a
bilinear form.\cite{FRE01} In other approaches, which satisfactorily
implement local correlations, long range Coulomb interactions are typically
difficult to handle. 

The flexibility of the slave boson approach has contributed to its success in
many fields from Kondo physics \cite{COL84, REA83} and Kondo/Anderson lattices
\cite{TES86, HAR90, PEP07} to interfaces of correlated electronic systems.
\cite{PAV06, RUE07} It separates in a straightforward way the high and low
energy scales. By now the respective mean-field evaluations are well
documented  (see, e.g., Refs.~[\onlinecite{LIL90,FRE91,YUA02,SEI98,RAC06}]),
and a number of fluctuation calculations have been performed,
\cite{BAN92,ZIM97,KOC01,ZLA03} even though the choice of the proper framework
has been intensely debated. \cite{JOL91,BAN92,FRE92,ARR94,FRE01} In the case
of the single impurity Anderson model (SIAM) one rather resorts to
diagrammatic approaches which include slave boson techniques (for a recent
reference see Ref.~[\onlinecite{KIR04}]).

It was doubted that the low energy physics is implemented consistently in the
mean field evaluations because the decomposition is a local scheme. In the
present work we want to access all energy scales for a specific model and
compare the saddle point evaluation of the KR slave boson approach to exact
results. Comparing exact and saddle point calculations analytically
represents a difficult task when handling a large system. Therefore, as a
first step, we restrict our considerations to a two-site cluster which can be
diagonalized exactly but still incorporates characteristics of a strongly
correlated electron system: the considered two site cluster presents a
truncated infinite-$U$ single-impurity Anderson model with a nearest neighbor
Coulomb interaction. One site is strongly correlated with an infinite local
Coulomb repulsion which hybridizes with the second site, on which the local
Coulomb repulsion vanishes. From the investigation of this model another
benchmark of the slave boson approach is provided here and, moreover, two
essential questions may be addressed: how do thermodynamics and correlation
functions compare between exact and saddle point evaluation? How is the
intersite Coulomb interaction appropriately introduced in the KR slave boson
approach?

{\it Interacting two-site cluster.}\,
We introduce the SIAM-type Hamiltonian

\begin{eqnarray}\label{eqh1}
{\mathcal H} & = & \sum_{\sigma} \left( \epsilon_{\rm c} c^{\dagger}_{\sigma}
  c^{\phantom{\dagger}}_{\sigma} + \epsilon_{\rm d} d^{\dagger}_{\sigma}
  d^{\phantom{\dagger}}_{\sigma} + V
  \left(c^{\dagger}_{\sigma}d^{\phantom{\dagger}}_{\sigma} + {\rm
  h.c.}\right)\right) \nonumber \\
& & +  U %\prod_{\sigma =\uparrow,\downarrow} 
d^{\dagger}_{\uparrow}  d^{\phantom{\dagger}}_{\uparrow}  
d^{\dagger}_{\downarrow}  d^{\phantom{\dagger}}_{\downarrow}  
+ I n_{\rm d}n_{\rm c}
\end{eqnarray}
which is defined on two sites. 
The operators $c^{\dagger}_{\sigma}$
($c^{\phantom{\dagger}}_{\sigma}$) and $d^{\dagger}_{\sigma}$
($d^{\phantom{\dagger}}_{\sigma}$) represent the creation
(annihilation) of  ``band'' and ``impurity'' electrons,
with spin projection $\sigma$. The energies $\epsilon_{\rm c}$ and
$\epsilon_{\rm d}$ are the band and impurity energy levels, respectively,
while the hybridization energy is $V$. Here $U$ is the  on-site repulsion,
which is the largest energy scale in the model and it will be set to infinity
as in the standard infinite-$U$ SIAM. Finally, 
${\mathcal H}_I \equiv I n_{\rm d}n_{\rm c}$ represents the non-local Coulomb
interaction. 

This extended two-site SIAM may be solved exactly, either 
by diagonalization of the Hamiltonian,  Eq.~(\ref{eqh1}), or, in the Lagrangian
language,\cite{FRE07,FRE08} through the exact evaluation of the path integrals
representing the desired quantities within 
a slave boson representation in the radial gauge.\cite{FRE01}
The two-particle basis for the Hamiltonian matrix consists of two singlet
states and three triplet states. Indeed one finds a threefold degenerate
eigenvalue of ${\mathcal H} $
\begin{equation}\label{lambdat}
E^{(t)} =  (\epsilon_{\rm
  c}+\epsilon_{\rm d} + I ) 
\end{equation}
corresponding to the triplet states, and two non-degenerate eigenvalues
controlled through $\Delta \equiv \epsilon_{\rm c} - \epsilon_{\rm d}$: 
\begin{eqnarray}\label{eigva}
E^{(s)}_{\pm} \!\!=\! \frac{1}{2}
  \left(3 \epsilon_{\rm c}+\epsilon_{\rm d} + I  \pm
  \sqrt{(\Delta - I)^2 + 8 V^2} \right)
\end{eqnarray}
corresponding to the singlet states.

{\it KR slave boson scheme.}\,
The goal is now to formulate an effective theory which correctly
reproduces the expectation value of the charge density and the charge and spin
density correlation functions. The requirement to work with canonical
fermionic or bosonic operators (or rather `fields' in the Lagrangian language)
leads to the slave boson scheme with which we intend to evaluate the free
energy in the KR SBMF (slave boson mean field or, equivalently, saddle point)
approximation. The first step in the scheme is to enlarge the Fock space for
the impurity site by introducing auxiliary fermions 
$f^{({\dagger})}_{\sigma}$ and slave bosons $e^{({\dagger})}$ and 
$p^{({\dagger})}_{\sigma}$, where $e^{({\dagger})}$, acts on an empty
and $p^{({\dagger})}_{\sigma}$ on a singly occupied impurity site. A field for
double occupancy of the impurity site is omitted as we restrict the evaluation
to infinite $U$. The electron operator on the impurity is then represented as
a composite operator 
\begin{equation}
\label{sb}
d^{\dagger}_{\sigma}=f^{\dagger}_{\sigma}p^{\dagger}_{\sigma}\,e\, ,
\qquad d^{\phantom{\dagger}}_{\sigma}=e^{\dagger}\,
p^{\phantom{\dagger}}_{\sigma}f^{\phantom{\dagger}}_{\sigma}\, .
\end{equation}
These fields are subject to three constraints:
\begin{eqnarray}
e^{\dagger} e &+& \sum_{\sigma} p^{\dagger}_{\sigma} p_{\sigma} = 1 \nonumber\\
p^{\dagger}_{\sigma} p_{\sigma} &=& f^{\dagger}_{\sigma} f_{\sigma} 
\quad \sigma = \uparrow,\;\downarrow
\end{eqnarray}
which are enforced by three Lagrange multipliers, denoted by $\alpha$ and
$\lambda_{\sigma}$, respectively. If strictly enforced, only empty and singly
occupied states in the physical space remain.  

Now, in the second step of the setup, the KR technique renormalizes the
coupling parameters of the Hamiltonian or Lagrangian. It thereby makes use of
the freedom to choose different (operator) representations of the respective
coupling terms, which are all equivalent in the physical subspace. This
freedom originates from the decomposition of the electron field into auxiliary
fields. Kotliar and Ruckenstein~\cite{KOT86} introduced $z$-factors into the
hybridization term which then reads for the SIAM: 
\begin{eqnarray}
\label{eqhV}
{\mathcal H_V} & = &  V \sum_{\sigma}
\left(c^{\dagger}_{\sigma}d^{\phantom{\dagger}}_{\sigma} + {\rm
  h.c.}\right)  \nonumber \\
& = &  V \sum_{\sigma}
\left(c^{\dagger}_{\sigma} z_{\sigma} f^{\phantom{\dagger}}_{\sigma} + {\rm
  h.c.}\right) \, .
\end{eqnarray}
A straightforward choice would be $z_\sigma = e^{\dagger}\,
p^{\phantom{\dagger}}_{\sigma}$ which directly translates the relation
Eq.~(\ref{sb}) within ${\mathcal H_V} $. However, the well established
substitution is rather 
\begin{equation}
\label{zfactor}
 z_\sigma =  e^{\dagger}\,  (1-p^{\dagger}_{\sigma} 
p^{\phantom{\dagger}}_{\sigma})^ {-\frac{1}{2}}\,
 (1- e^{\dagger}e - p^{\dagger}_{-\sigma} 
p ^{\phantom{\dagger}}_{-\sigma})^{-\frac{1}{2}} 
 \, p^{\phantom{\dagger}}_{\sigma}
\end{equation}
which reproduces the $U=0$ spinless case and the Gutzwiller
approximation~\cite{KOT86} in the saddle point evaluation and which led to the
excellent agreement of the SBMF with numerical simulations in the case of the
single band Hubbard model. We note that ${\mathcal H_V} $ has the same matrix
elements in the physical subspace for any of these representations 
for $z_\sigma$. 

In the third step of the scheme the bosonic fields, including the Lagrange
multipliers, are replaced by their respective saddle point values. Here we
emphasize that $z_\sigma^2$ is in fact the quasiparticle weight, as may be
confirmed by straightforward evaluation of
$\frac{\partial\Sigma(\omega)}{\partial\omega} = 1- \frac{1}{z_\sigma^2}$,
where $\Sigma(\omega)$ represents the impurity site self-energy. 

Before we discuss the KR SBMF results, the renormalization of
the nonlocal Coulomb interaction $I$ has to be introduced: 
\begin{equation}
\label{eqhI}
{\mathcal H}_I = I n_{\rm d}n_{\rm c}
 =  I\,  \sum_{\sigma} \bigl(1-y^{\dagger}_{\sigma} y_{\sigma}\bigr)\, n_{{\rm
 c},\sigma} \, .
\end{equation}
The natural choice would be $y_\sigma = e$ (spin independent), as
$(1-e^\dagger e)$ is the density operator. However, on the saddle point level,
we rather expect that  $y_\sigma=z^\dagger_\sigma$ should account for a more
appropriate representation. As $z^\dagger_\sigma z_\sigma$ is related to the
quasiparticle weight, $1 - y^\dagger_\sigma y_\sigma$ characterizes the
incoherent part of an electronic excitation. It here refers to a local
process, Eq.~(\ref{eqhI}). Correspondingly we define: 
\begin{equation}
\label{yfactor}
 y^\dagger_\sigma =
  e^{\dagger}\,
 (1+\varepsilon-p^{\dagger}_{\sigma} 
p^{\phantom{\dagger}}_{\sigma})^ {-\frac{1}{2}}\,
 (1- e^{\dagger}e - p^{\dagger}_{-\sigma} 
p ^{\phantom{\dagger}}_{-\sigma})^{-\frac{1}{2}} 
  \, p^{\phantom{\dagger}}_{\sigma} \, .
\end{equation}
An infinitesimal convergence factor $\varepsilon > 0$ has been introduced to
ensure the property $y_\sigma= 0$ for $e = 0$, also on the saddle point
level. This singular assignment of $y_\sigma= 0$ may occur if the
hybridization is suppressed either due to the formation of a triplet state or
on account of the special choice $V=0$. Otherwise $\varepsilon$ may be set to
zero from the outset. Such a convergence factor is not necessary 
for $z_\sigma$ in the hybridization, Eq.~(\ref{eqhV}).
We emphasize again that replacing $ I n_{\rm d}n_{\rm c}$ by the expression on
the rhs of Eq.~(\ref{eqhI}) with $y^\dagger_\sigma$ from Eq.~(\ref{yfactor})
does not hurt the correctness of the representation, i.e., all matrix elements
of ${\mathcal H}_I$ are unaffected in the physical subspace. Moreover, by
extending the scheme developped in Refs.~[\onlinecite{FRE07,FRE08}] to the KR
representation, it can be shown by direct evaluation of the path integral that
the exact partition function is recovered when using Eqs.~(\ref{eqhI},
\ref{yfactor}). 

{\it KR SBMF.}\,
The saddle point evaluation is now easily implemented: (i) the bosonic fields
are replaced 
by  real variables and (ii) these variables are found from the minimization of
the grand canonical potential. With (i) we identify the fermionic matrix as: 
\begin{equation}\label{fmatrix}
\left[{\mathcal E}_{\sigma}\right]=
\left(\begin{array}{cc}
E_{{\rm c},\sigma} & z_\sigma V\\
z_\sigma V & E_{{\rm f},\sigma}\\
\end{array}\right).
\end{equation}
with $E_{{\rm c},\sigma} = \epsilon_{\rm c} + (1-y_\sigma^2)\,I  -\mu $ and 
$E_{{\rm f},\sigma} = \epsilon_{\rm d} +\lambda_\sigma  -\mu $. Its
eigenvalues are 
\begin{equation}
\label{KREV}
 E_{\pm,\sigma} = \frac{1}{2}\bigl(E_{{\rm c},\sigma} + E_{{\rm f},\sigma}  
 \pm \sqrt{(E_{{\rm c},\sigma} - E_{{\rm f},\sigma})^2 + 4 z_\sigma^2 V^2}
 \; \bigr)
\end{equation}
For convenience, we have introduced a joint chemical potential $\mu$ for all
electrons (on the $d$- and the $c$-site) to control the filling. 

Part (ii) of the evaluation depends crucially on the number of electrons $N$
is the system. We first 
consider the two-electron case ($N=2$). The free energy $F$ reads: 
$F(N=2,T) = - T\sum_{\sigma,r=\pm} 
\ln (1+\exp({-\beta E_{r,\sigma}})) - \sum_\sigma \lambda_\sigma p_\sigma^2 -
\alpha (1-e^2-\sum_\sigma p_\sigma^2) + 2\mu$ where the first log-term is
standard and the terms with the Lagrange multipliers $\lambda_\sigma$ and
$\alpha$ were generated through the constraints. The phases of the slave boson
fields do not enter into the saddle point evaluation, 
so $e$ and $p_\sigma$ are now real numbers, as well as $\alpha$ and
$\lambda_\sigma$. The quadratic term $e^2$ is to be identified with the hole
number $x$ on the correlated site, i.e., $x=e^2$. We take the temperature $T$
in units of $k_B$ and  $\beta = 1/T$. For the limit $T\rightarrow 0$ we have
$F(N=2,0)=\sum_\sigma  E_{-,\sigma} - \sum_\sigma \lambda_\sigma p_\sigma^2 
+ \alpha (e^2 + \sum_\sigma p_\sigma^2 -1) + 2\mu$. The minimization of the
ground state $F(N=2,0)$ is cumbersome but straightforward. We obtain: 
\begin{eqnarray}
\label{Feq}
\!\!\! \!\!\!  F (N = 2\!\! \! \!\!\!&&,T=0) - 2\epsilon_{\rm d}= \nonumber  \\
&&  \!\!\!  \!\!\!   \Bigl(
\Delta(1+x) + I (1-x) -2V\sqrt{2x}\sqrt{1-x}
\Bigr)
\end{eqnarray}
Here,  the hole number on the correlated site obeys the analytical relation
(with $\nu=-1$ and $x_{-}=x$): 
\begin{equation}
\label{xeq}
x_{\nu} = \frac{1}{2}\biggl(1 + \nu\;\frac{\Delta -I}
{ \sqrt{(\Delta-I)^2+8 V^2}} \biggr)\; ,
\end{equation}
which displays a  monotonous decay with increasing  $(\Delta -I)/V$. The
ground state is paramagnetic and, with $p_\sigma^2 = p_{-\sigma}^2 \equiv
p^2$, the constraint yields the correct relation $p^2= (1-x)/2$ for the
expectation value of single occupancy on the correlated site. Notably, these
relations, Eqs.~(\ref{Feq},\ref{xeq}), are exact. This statement is consistent
with the observation that the lowest two-particle state acquires the energy
$(E_{-,\uparrow}+ E_{-,\downarrow})= E^{(s)}_{-}$ of the lower
antiferromagnetic singlet,
Eq.~(\ref{eigva}), for the saddle point solution of the bosonic fields. 

However, it is not only the lowest energy state in the two particle sector,
which is solved exactly in the KR SBMF scheme. The approach also provides the
exact solutions for the single particle $ (N=1)$ and the three particle $
(N=3)$ sectors, and moreover the triplet solution for $ (N=2)$. These latter
cases are not as surprising as the two-particle solution for the paramagnetic
ground state. Nevertheless, these states will be important for the calculation
of correlation functions. 

Here it suffices to emphasize that the KR approach was set up with the proviso
that the spin polarized limit is correctly implemented through the weight
factors $z_\sigma$.  \cite{KOT86} In fact, we find for $N=1$ the hole
expectation value on the correlated site $x = \frac{1}{2} \bigl(1 - \Delta/ 
\sqrt{\Delta^2+4 V^2} \bigr)$ and the fermionic matrix Eq.~(\ref{fmatrix}) 
reduces to the simple $2\times 2$ matrix with entries $\epsilon_{\rm c}$ and
$\epsilon_{\rm d}$ on the diagonal and $V$ as the off-diagonal matrix
elements. This represents correctly a two-level problem. The $N=2$ triplet
state and the $N=3$ case are exactly reproduced because these are local states
with vanishing kinetic term ${\mathcal H_V}$ and the weight factors $y_\sigma$
for the nearest neighbor interaction $I$ have been chosen suitably in
Eq.~(\ref{yfactor}). Truly, for the triplet state, we could verify that 
$(E_{-,\uparrow}+ E_{+,\uparrow})= E^{(t)}$ for $E_{\pm,\uparrow}$ from the 
corresponding saddle point solution. There, the solution becomes the ground
state if the model free energy is  extended by a magnetic term, linear in the
external field. 

The high energy singlet state presents, for the limit $V\rightarrow 0$, a
local singlet with the two electrons on the uncorrelated  site. For $V\neq 0$,
this singlet becomes non-local as the electronic states hybridize with the
correlated site. It is important to observe that the KR SBMF scheme is
flexible enough to accommodate also this second non-local two-particle state 
correctly. In order to gain the exact expression  $E^{(s)}_{+}$ of
Eq.~(\ref{eigva}), the saddle point evaluation for the bosonic fields has to
be retraced, however, with $(E_{+,\uparrow}+ E_{+,\downarrow})$ for the
two-particle state. We find for the hole number on the correlated site
$x=x_{+}$ (see Eq.~(\ref{xeq})) and $(E_{+,\uparrow}+ E_{+,\downarrow})=
E^{(s)}_{+}$. In other words, we here minimize the energy in the subspace of
the Hilbert space that is orthogonal to the ground state.

{\it Spin Correlation function.}\, 
The spin correlation function involves transitions between states in the
singlet and triplet spin sector which are separated by $E^{(t)}- E^{(s)}_{-}$.
For the two-site model we can easily evaluate the spin correlation function 
$\langle \phi^{(s)}_{\pm}| S_{+}(t) S_{-}(0)|\phi^{(s)}_{\pm}\rangle$
where $\phi^{(s)}_{\pm}$ refers to the singlet states and $S_{-}(0) = 
d^{\dagger}_{\downarrow}d^{\phantom{\dagger}}_{\uparrow}=  S^{\dagger}_{+}(0)$
acts on the correlated (impurity) site. The singlet state ket-vector is:
\begin{equation}\label{singletket}
|\phi^{(s)}_{\pm}\rangle  = 
    a\; (c^{\dagger}_{\uparrow} c^{\dagger}_{\downarrow})|0\rangle
    \pm
    b \; 
     (d^{\dagger}_{\uparrow} c^{\dagger}_{\downarrow} + 
      c^{\dagger}_{\uparrow} d^{\dagger}_{\downarrow})|0\rangle
\end{equation}
where the fermionic operators act on their vacuum state $|0\rangle$, and  the
normalization factors are $a=\sqrt{x } $ and 
$b= \sqrt{\frac{1}{2}(1- x)}$. In fact, the total 
spin operator $ S_{+/-}^{({\rm tot})}$  for the two sites correctly yields the
singlet state property $ S_{+/-}^{({\rm tot})}|\phi^{(s)}_{\pm}\rangle = 0 $. 
The spin correlation function for the correlated site is, e.g., 
for $T\ll E^{(t)}- E^{(s)}_{-}$ and $\Delta-I>0$, and for a chemical
potential which fixes the particle number to two:
\begin{equation}
   \langle \phi^{(s)}_{-}| e^{-\beta {\mathcal H}} 
    e^{-it {\mathcal H}}
    d^{\phantom{\dagger}}_{\uparrow} d^{\dagger}_{\downarrow}
    e^{it {\mathcal H}}
    d^{\phantom{\dagger}}_{\downarrow}d^{{\dagger}}_{\uparrow} 
     |\phi^{(s)}_{-}\rangle \nonumber 
\end{equation}
The procedure for its evaluation in the KR technique is straightforward: the
exponential Hamiltonian terms are taken in their respective mean field form
which act on the states generated by the spin operators. As the saddle point
evaluation delivers the exact energy levels, we identify: 
\begin{equation}\label{spincorrel}
\langle S_{+}(t) S_{-}(0)\rangle =\frac{1}{2}(1- x ) \, e^{it (E^{(t)}- E^{(s)}_{-})} 
\end{equation}
where the partition function $Z=e^{-\beta  E^{(s)}_{-} } $ cancels the
temperature factor. Similarly, the exact spin correlation function with a
trace, which includes also the triplet and high energy singlet, is recovered
for arbitrary temperature values. 

It is the remarkable flexibility of the slave boson technique which correctly
renders certain limits of  strongly correlated electron systems. This
adaptability  originates in the freedom to choose appropriate 
weight factors for the interaction terms in the projected field theory. 
For the two-site SIAM, the generic choice of this weight factors already leads
to the exact solution of the model. The choice is generic in the sense that it
was introduced early on for the Hubbard model in the original work on the KR
slave boson scheme. \cite{KOT86} It is striking to observe that even the
inclusion of a non-local Coulomb repulsion in this model does not break the 
exactitude of the saddle point evaluation in this scheme.

%%%%%%%%%%%%%%%%%%%%%%%%%%%%%%%%%%%%%%%%%%%%%%

{\it Acknowledgments.}\, 
Illuminating discussions with P.~W\"olfle are gratefully acknowledged. The
work was supported by DFG through SFB 484. T.~K.\ is grateful for the kind
hospitality at the CRISMAT in Caen and at the Aspen Center for Physics where
part of this work has been accomplished.

\end{document}